# Histograms and Wavelets on Probabilistic Data

Graham Cormode[*]      Minos Garofalakis[†]

November 6, 2018

**Abstract**

There is a growing realization that uncertain information is a first-class citizen in modern database management. As such, we need techniques to correctly and efficiently process uncertain data in database systems. In particular, data reduction techniques that can produce concise, accurate synopses of large probabilistic relations are crucial. Similar to their deterministic relation counterparts, such compact probabilistic data synopses can form the foundation for human understanding and interactive data exploration, probabilistic query planning and optimization, and fast approximate query processing in probabilistic database systems.

In this paper, we introduce definitions and algorithms for building histogram- and wavelet-based synopses on probabilistic data. The core problem is to choose a set of histogram bucket boundaries or wavelet coefficients to optimize the accuracy of the approximate representation of a collection of probabilistic tuples under a given error metric. For a variety of different error metrics, we devise efficient algorithms that construct optimal or near optimal $B$-term histogram and wavelet synopses. This requires careful analysis of the structure of the probability distributions, and novel extensions of known dynamic-programming-based techniques for the deterministic domain. Our experiments show that this approach clearly outperforms simple ideas, such as building summaries for samples drawn from the data distribution, while taking equal or less time.

## 1 Introduction

Modern real-world applications generate massive amounts of data that is often *uncertain and imprecise*. For instance, data integration and record linkage tools can produce distinct degrees of confidence for output data tuples (based on the quality of the match for the underlying entities) [7]; similarly, pervasive multi-sensor computing applications need to routinely handle noisy sensor/RFID readings [22]. Motivated by these new application requirements, recent research efforts on *probabilistic data management* aim to incorporate uncertainty and probabilistic information as "first-class citizens" of the database system.

Among different approaches for modeling uncertain data, *tuple- and attribute-level uncertainty models* have seen wide adoption both in research papers as well as early system prototypes [1, 2, 3]. In such models, the attribute values for a data tuple are specified using a probability distribution over different mutually-exclusive alternatives (that might also include *non-existence*, i.e., the tuple is not present in the data), and assuming independence across tuples. The popularity of tuple-/attribute-level uncertainty models is due to both their simplicity of representation in current relational systems, as well as their intuitive query semantics. In essence, a probabilistic database is a concise representation for a probability distribution over an exponentially-large collection of *possible worlds*, each representing a possible "grounded" (deterministic) instance of the database (e.g., by flipping appropriately-biased independent coins to select an instantiation for each uncertain tuple). This "possible-worlds" semantics also implies clean semantics for queries over a probabilistic database — essentially, the result of a probabilistic query defines a probability distribution over the space of possible query results across all possible worlds [7]. The goal of query answering over uncertain data is to provide the expected value of the answer, or tail bounds on the distribution of answers.

---

[*]AT&T Labs–Research, graham@research.att.com

[†]Yahoo! Research and UC Berkeley, minos@yahoo-inc.com

Unfortunately, despite its intuitive semantics, the paradigm shift towards tuple-level uncertainty also implies computationally-intractable *#P-hard data complexity* even for simple query processing operations [7]. These negative complexity results for query evaluation raise some serious practicality concerns for the applicability of probabilistic database systems in realistic settings. One possible avenue of attack is through the use of *approximate query processing* techniques over probabilistic data. As in conventional database systems, such techniques need to rely on effective *data reduction methods* that can effectively compress large amounts of data down to concise *data synopses* while retaining the key statistical traits of the original data collection [9]. It is then feasible to run more expensive algorithms over the much compressed representation, and still obtain a fast and accurate answer. In addition to enabling fast approximate query answers over probabilistic data, such compact synopses can also provide the foundation for human understanding and interactive data exploration, and probabilistic query planning and optimization.

The data-reduction problem for deterministic databases is well understood, and several different synopsis construction tools exist. *Histograms* [17] and *wavelets* [4] are two of the most popular general-purpose data-reduction methods for conventional data distributions. It is therefore meaningful and interesting to build histogram and wavelet synopses of probabilistic data. Here, histograms, as on traditional "deterministic" data, divide the given input data into "buckets" so that all tuples falling in the same bucket have similar behavior: the bucket boundaries are chosen to minimize a given error function that measures this within-bucket dissimilarity. Likewise, wavelets represent the probabilistic data by choosing a small number of wavelet basis functions which best describe the data, and contain as much of the "expected energy" of the data as possible. So for both histograms and wavelets, the synopses aim to capture and *describe* the probabilistic data as accurately as possible given a fixed size of synopsis. This description is clearly of use both to present to users to compactly show the key components of the data, as well as in approximate query answering and query planning. As in the large body of work on synopses for deterministic data, we consider a number of standard error functions, and show how to find the optimal synopsis relative to this class of function.

There has been much recent work that has explored many different variants of the basic synopsis problem, and we summarize the main contributions in Section 2. Unfortunately, this work is not applicable when moving to *probabilistic* data collections that essentially represent a huge set of possible data distributions (i.e., "possible worlds"). It may seem that probabilistic data can be thought of defining a weighted instance of the deterministic problem, but we show empirically and analytically that this yields poor representations. Thus, building effective histogram or wavelet synopses for such collections of possible-world data distributions is a challenging problem that mandates novel analysis and algorithmic approaches.

**Our Contributions.** In this paper, we provide the first formal definitions and algorithmic solutions for constructing histogram and wavelet synopses over probabilistic data. In particular:

- We define the "probabilistic data reduction problem" over a variety of cumulative and maximum error objectives, based on natural generalizations of histograms and wavelets from deterministic to probabilistic data.

- For histograms, we give efficient techniques to find the optimal histogram under all common cumula-

tive error objectives (such as sum-squared error and sum-relative error), and corresponding maximum error objectives. We also show that fast approximate solutions are possible. Our techniques rely on careful analysis of each objective function in turn, and showing that the cost of a given bucket, along with its optimal representative value, can be found efficiently from appropriate precomputed arrays. These results require careful proof, due to the distribution of values each item can take on, and the potential correlations between items.

- For wavelets, we similarly show optimal techniques for the core sum-squared error (SSE) objective. Here, it suffices to compute the wavelet transformation of a deterministic input derived from the probabilistic input. We also show how to extend algorithms from the deterministic setting to probabilistic data for non-SSE objectives.

- We report on experimental evaluation of our methods, and show that they achieve appreciably better results than simple heuristics. The space and time costs are always equal or better than the heuristics.

After surveying prior work on uncertain data, we describe the relevant data models and synopsis objectives in Section 2. Our results on Histograms and Wavelets are in Sections 3 and 4, respectively. We show experimental analysis of our techniques in Section 5, and then suggest some future directions.

## 1.1 Prior Work on Uncertain Data

Key ideas in probabilistic databases are presented in tutorials by Dalvi and Suciu [24, 8], and built on by systems such as Trio [2] and MayBMS [1]. Initial research has focused on how to store and process uncertain data within database systems, and hence how to answer SQL-style queries. Subsequently, there has been a growing realization that in addition to storing and processing uncertain data, there is a need to run advanced algorithms to analyze and mine uncertain data. Recent work has analyzed how to compute properties of streams of uncertain tuples such as the expected average and number of distinct items [5, 20, 21]; how to cluster uncertain data [6]; and finding frequent items within uncertain data [26].

To the best of our knowledge, no prior work has studied problems of building histogram and wavelet synopses of probabilistic data. There has been work on finding quantiles of unidimensional data [5, 21], which can be thought of as the *equi-depth* histogram; the techniques to find these show that it simplifies to the problem of finding quantiles over weighted data, where the weight of each item is simply its expected frequency. Similarly, finding frequent items is somewhat related to finding high-biased histograms. Lastly, we can also think of building a histogram as being a kind of clustering of the data along the domain. However, the nature of the error objectives on histograms that are induced by the formalizations of clustering are quite different from here, and so probabilistic clustering techniques [6] do not give good solutions for histograms.

## 2 Preliminaries and Problem Formulation

## 2.1 Probabilistic Data Models

A variety of models of probabilistic data have been proposed. The different models capture various levels of independence between the individual data values described. Each model describes a distribution over *possible worlds*: each possible world is a (traditional) relation containing some number of tuples. The most general model describes the complete correlations between all tuples; effectively, it describes every possible world and its associated probability explicitly. However, the size of such a model for even a moderate number of tuples is immense, since the exponentially many possible combinations of values are spelled out. In practice, finding the (exponentially many) parameters for the fully general model is unfeasible; instead, more compact models are adopted which can reduce the number of parameters by making independence assumptions between tuples.

The simplest probabilistic model is the *basic* model, which consists of a sequence of tuples containing a single value and the probability that it exists in the data. More formally,

**Definition 1.** *The* basic model *consists of a set of $m$ tuples where the $j$th tuple consists of a pair $\langle t_j, p_j \rangle$. Here, $t_j$ is an item drawn from a fixed domain, and $p_j$ is the probability that $t_j$ appears in any possible world. Each possible world $W$ is formed by the inclusion of a subset of the items $t_j$. We can write $j \in W$ if $t_j$ is present in the possible world $W$, and $j \notin W$ otherwise. Each tuple is assumed to be independent of all others, so the probability of a possible world $W$ is given by*

$$\Pr[W] = \prod_{j \in W} p_j \prod_{j \notin W} (1 - p_j).$$

∎

Note here that the items $t_j$ can be somewhat complex (e.g. a row in a table), but without loss of generality we will treat them as simple objects. In particular, we will be interested in cases that can be modeled as when the $t_j$s are drawn from a fixed, ordered domain (such as $1, 2, 3 \ldots$) of size $n$, and several $t_j$s can correspond to occurrences of the same item. We consider two extensions of the basic model, which each capture dependencies not expressible in the basic model by providing a compact discrete probability density function (pdf) in place of the single probability.

**Definition 2.** *In the* tuple pdf *model, instead of a single (item, probability) pair, there is a set of pairs with probabilities summing to at most 1. That is, the input consists of a sequence of tuples $t_j \in \mathcal{T}$ of the form $\langle (t_{j1}, p_{j1}), \ldots (t_{j\ell}, p_{j\ell}) \rangle$. The interpretation is that each tuple specifies a set of mutually exclusive possible values for the $i$th row of a relation. We require that the sum of the probabilities within a tuple is at most 1; if less than unity, then the remaining probability measures the chance that there is no corresponding item. We can interpret this as describing a discrete pdf for the $j$th item in the input as $\Pr[t_j = t_{j1}] = p_{j1}$, and so on. Each tuple is assumed to be independent of all others, so that the probability of any possible world can be computed via careful multiplication of the relevant probabilities.* ∎

This model has been widely adopted, and is used in the TRIO [2] work and elsewhere [21]. It captures the case where an observer makes readings, and has some uncertainty over what was seen. An alternate case is when an observer makes readings of a known item (for example, this could be a sensor making discrete readings), but has uncertainty over a value or frequency associated with the item:

**Definition 3.** *The* value pdf *model consists of a sequence of tuples of the form $\langle i : (f_{i1}, p_{i1}) \ldots (f_{i\ell}, p_{i\ell}) \rangle$, where the probabilities in each tuple sum to at most 1. The interpretation is that tuple specifies the distribution of frequencies of a separate item; the distributions of different items are assumed to be distinct. We can interpret this as describing a discrete pdf for the random variable $g_i$ giving (say) the distribution of the frequencies of the $i$th item: $\Pr[g_i = f_{i1}] = p_{i1}$, and so on. Due to the independence, the probability of any possible world is computed via multiplication of probabilities for the frequency of each item in turn. If probabilities in a tuple sum to less than one, the remainder is taken to implicitly specify the probability that the frequency is zero, for easy compatibility with the basic model. Let the set of all values of frequencies used be $\mathcal{V}$, so every $(f, p)$ pair has $f \in \mathcal{V}$.* ∎

For both the basic and tuple models, the frequency of any given item $i$ within a possible world, $g_i$, is a non-negative integer, and each occurrence corresponds to a tuple from the input. The value pdf model can specify arbitrary fractional frequencies, but the number of such frequencies is bounded by the size of the input, $m$. Note that the basic model is a special case of the tuple pdf and value pdf model, but that neither of these two is contained within the other. However, input in the tuple pdf *induces* a distribution over frequencies of each item, so we can define the *induced value pdf* which, for each item $i$, provides $\Pr[g_i = v]$ for some $v \in \mathcal{V}$. The important detail is that, unlike in the value pdf model, these induced pdfs are *not* independent; nevertheless, this representation is useful in our subsequent analysis. For data presented in the tuple pdf format, it is straightforward to build induced the value pdf for each value inductively, taking time $O(|\mathcal{V}|)$ to update the value pdf built so far. The space required is linear in the size of the input, $O(m)$.

**Definition 4.** *Given an input in any of these models, let $\mathcal{W}$ denote the space of all possible worlds, and $\Pr[W]$ denote the probability associated with possible world $W \in \mathcal{W}$. We can then compute the* expectation *of various quantities over possible worlds as, given a function $f$ which can be evaluated on a possible world $W$,*

$$\mathsf{E}_{\mathcal{W}}[f] = \sum_{W \in \mathcal{W}} \Pr[W] f(W) \tag{1}$$

**Example 1.** *Consider the ordered domain containing the three items $1, 2, 3$. The input $\langle 1, \frac{1}{2} \rangle, \langle 2, \frac{1}{3} \rangle, \langle 2, \frac{1}{4} \rangle, \langle 3, \frac{1}{2} \rangle$ in the basic model defines the following twelve possible worlds and corresponding probabilities:*

| $W$ | $\emptyset$ | 1 | 12 | 122 | 123 | 1223 | 13 | 2 | 22 | 23 | 223 | 3 |
|---|---|---|---|---|---|---|---|---|---|---|---|---|
| $\Pr[W]$ | $\frac{1}{8}$ | $\frac{1}{8}$ | $\frac{5}{48}$ | $\frac{1}{48}$ | $\frac{5}{48}$ | $\frac{1}{48}$ | $\frac{1}{8}$ | $\frac{5}{48}$ | $\frac{1}{48}$ | $\frac{5}{48}$ | $\frac{1}{48}$ | $\frac{1}{8}$ |

*The input $\langle (1, \frac{1}{2}), (2, \frac{1}{3}) \rangle, \langle (2, \frac{1}{4}), (3, \frac{1}{2}) \rangle$ in the tuple pdf model defines the following eight possible worlds and corresponding probabilities:*

| $W$ | $\emptyset$ | 1 | 2 | 3 | 12 | 13 | 22 | 23 |
|---|---|---|---|---|---|---|---|---|
| $\Pr[W]$ | $\frac{1}{24}$ | $\frac{1}{8}$ | $\frac{1}{8}$ | $\frac{1}{12}$ | $\frac{1}{8}$ | $\frac{1}{4}$ | $\frac{1}{12}$ | $\frac{1}{6}$ |

*The input $\langle 1 : (1, \frac{1}{2}) \rangle, \langle 2 : (1, \frac{1}{3}), (2, \frac{1}{4}) \rangle, \langle 3 : (1, \frac{1}{2}) \rangle$ in the value pdf model defines the pdfs*

$$\Pr[g_1 = 0] = \tfrac{1}{2}, \Pr[g_1 = 1] = \tfrac{1}{2}$$
$$\Pr[g_2 = 0] = \tfrac{5}{12}, \Pr[g_2 = 1] = \tfrac{1}{3}, \Pr[g_2 = 2] = \tfrac{1}{4}$$
$$\Pr[g_3 = 0] = \tfrac{1}{2}, \Pr[g_3 = 1] = \tfrac{1}{2}$$

*and hence provides the following distribution over twelve possible worlds:*

| $W$ | $\emptyset$ | 1 | 12 | 122 | 123 | 1223 | 13 | 2 | 22 | 23 | 223 | 3 |
|---|---|---|---|---|---|---|---|---|---|---|---|---|
| $\Pr[W]$ | $\frac{5}{48}$ | $\frac{5}{48}$ | $\frac{1}{12}$ | $\frac{1}{16}$ | $\frac{1}{12}$ | $\frac{1}{16}$ | $\frac{5}{48}$ | $\frac{1}{12}$ | $\frac{1}{16}$ | $\frac{1}{12}$ | $\frac{1}{16}$ | $\frac{5}{48}$ |

*In all three cases, $\mathsf{E}_{\mathcal{W}}[g_1] = \frac{1}{2}$. In the value pdf case, $\mathsf{E}_{\mathcal{W}}[g_2] = \frac{5}{6}$, for the other two cases $\mathsf{E}_{\mathcal{W}}[g_2] = \frac{7}{12}$.*

Although two possible worlds may be formed in different ways, they may be indistinguishable: for instance, in the basic model example, the world $W = \{2\}$ can result either from the second or third tuple. We will typically not distinguish possible worlds based on how they arose, and so treat them as identical.

Our input is then characterized by parameters $n$, giving the size of the ordered domain from which the input is drawn, $m$, the total number of pairs in the input (hence the input can be described with $O(m)$ pieces of information), and $\mathcal{V}$, the set of values that the frequencies take on. Here $|\mathcal{V}| \leq m$, but could be much less. In the all three examples above, we have $n = 3, m = 4$, and $\mathcal{V} = \{0, 1, 2\}$.

## 2.2 Histogram and Wavelet Synopses

We review key techniques for synopses on deterministic data.

**Histograms on Deterministic Data.** Consider a one-dimensional data distribution defined (without loss of generality) over the integer domain $[n] = \{0, \ldots, n-1\}$. For each $i \in [n]$, we let $g_i$ denote the *frequency* of domain value $i$ in the underlying data set A *histogram* synopsis provides a concise, piece-wise approximate representation of the distribution based on partitioning the ordered domain $[n]$ into $B$ *buckets*. Each bucket $b_k$ consists of a *start and end point*, $b_k = (s_k, e_k)$, and approximates the frequencies of the contiguous subsequence of values $\{s_k, s_{k+1}, \ldots, e_k\}$ (termed the *span* of the bucket) using a single representative value $\hat{b}_k$. We also let $n_k$ ($e_k - s_k + 1$) denote the width (i.e., number of distinct items) of bucket $b_k$. The $B$ buckets in a histogram form a *partition* of $[n]$; that is, $s_1 = 0$, $e_B = n-1$, and $s_{k+1} = e_k + 1$ for all $k = 1, \ldots, B-1$.

By using $O(B) \ll n$ space to represent an $O(n)$-size data distribution, histograms provide a very effective means of data reduction, with numerous applications [9]. This data reduction also implies *approximation errors* in the estimation of frequencies, since each $g_i \in b_k$ is estimated as $\hat{g}_i = \hat{b}_k$. The histogram construction problem is, given a storage budget $B$, build a $B$-bucket histogram $\mathcal{H}_B$ that is *optimal under some aggregate error metric*. Important histogram error metrics to minimize include, for instance, the *Sum-Squared-Error*

$$\text{SSE}(\mathcal{H}) = \sum_{i=1}^{n}(g_i - \hat{g}_i)^2 = \sum_{k=1}^{B}\sum_{i=s_k}^{e_k}(g_i - \hat{b}_k)^2$$

(which defines the important class of *V-optimal* histograms [18, 19]) and the *Sum-Squared-Relative-Error*:

$$\text{SSRE}(\mathcal{H}) = \sum_{i=1}^{n} \frac{(g_i - \hat{g}_i)^2}{\max\{c, |g_i|\}^2}$$

(where the sanity-bound constant $c$ in the denominator is used to avoid excessive emphasis being placed on small frequency values [10, 16]). The *Sum-Absolute-Error* (SAE) and *Sum-Absolute-Relative-Error* (SARE) are defined similarly to SSE and SSRE, replacing the square with an absolute value so that

$$\text{SAE}(\mathcal{H}) = \sum_{i=1}^{n} |g_i - \hat{g}_i| \text{ and } \text{SARE}(\mathcal{H}) = \sum_{i=1}^{n} \frac{|g_i - \hat{g}_i|}{\max\{c, |g_i|\}}$$

In addition to such *cumulative* metrics, *maximum error metrics* have also been employed in order to provide approximation guarantees on the relative/absolute error of individual frequency approximations [16]; these include, for example, *Maximum-Absolute-Relative-Error* $\text{MARE}(\mathcal{H}) = \max_{i \in [n]} \frac{|g_i - \hat{g}_i|}{\max\{c, |g_i|\}}$.

Histogram construction satisfies the *principle of optimality:* If the $B^{th}$ bucket in the optimal histogram spans the range $[i+1, n-1]$, then the remaining $B-1$ buckets must form an optimal histogram for the range $[0, i]$. This immediately leads to a Dynamic-Programming (DP) algorithm for computing the optimal error value $\text{OPTH}[j, b]$ for a $b$-bucket histogram spanning the prefix $[1, j]$ based on the following recurrence:

$$\text{OPTH}[j, b] = \min_{0 \leq i < j}\{h(\text{OPTH}[l, b-1], \min_{\hat{b}}\{\text{BERR}([l+1, j], \hat{b})\})\}, \quad (2)$$

where $\text{BERR}([x, y], z)$ denotes the error contribution of a *single* histogram bucket spanning $[x, y]$ using a representative value of $z$ to approximate all enclosed frequencies, and $h(x, y)$ is simply $x + y$ (respectively, $\max\{x, y\}$) for cumulative (resp., maximum) error objectives. The key to translating the above recurrence into a fast histogram construction algorithm lies in being able to quickly find the best representative $\hat{b}$ and the corresponding optimal error value $\text{BERR}()$ for the *single-bucket case*.

For example, in the case of the SSE objective the representative value minimizing a bucket's SSE contribution is exactly the average bucket frequency $\hat{b}_k = \frac{\sum_{i=s_k}^{e_k} g_i}{n_k}$ giving an optimal bucket SSE contribution of:

$$\min_{\hat{b}_k}\{\text{BERR}(b_k = [s_k, e_k], \hat{b}_k)\} = \sum_{i=s_k}^{e_k} g_i^2 - \frac{1}{n_k}\Big(\sum_{i=s_k}^{e_k} g_i\Big)^2.$$

By precomputing two $n$-vectors that store the prefix sums $\sum_{i=0}^{j} g_i$ and $\sum_{i=0}^{j} g_i^2$ for each $j = 0, \ldots, n-1$, the SSE contribution for any bucket $b_k$ in the DP recurrence can be computed in $O(1)$ time, giving rise to an $O(n^2 B)$ algorithm for building the SSE-optimal (or, V-optimal) $B$-bucket histogram [19]. Similar ideas apply for the other error metrics discussed above. For instance, the optimal MARE contribution for a single bucket depends only on the maximum/minimum frequencies. Using appropriate precomputed data structures on dyadic ranges leads to an efficient, $O(n \log^2 nB)$ DP algorithm for building MARE-optimal histograms [16].

**Wavelets on Deterministic Data.** *Haar wavelet synopses* [4, 11, 12, 25] provide another data reduction tool based on the Haar Discrete Wavelet Decomposition (DWT) [23] for hierarchically decomposing functions. At a high level, the Haar DWT of a data distribution over $[n]$ consists of a coarse overall approximation (the average of all frequencies) together with $n - 1$ *detail coefficients* (constructed through recursive pairwise averaging and differencing) that influence the reconstruction of frequency values at different scales. The Haar DWT process can be visualized through a binary *coefficient tree* structure, like the one depicted in Figure 1 for an example frequency array $A = [2, 2, 0, 2, 3, 5, 4, 4]$: Leaf nodes $g_i$ correspond to the original data distribution values in $A[]$. The root node $c_0$ is the overall average frequency, whereas each internal node $c_i$ ($i = 1, \ldots, 7$) is a detail coefficient computed as the half the difference between the average of frequencies in $c_i$'s left child subtree and the average of frequencies in $c_i$'s left child subtree (e.g., $c_3 = \frac{1}{2}(\frac{3+5}{2} - \frac{4+4}{2}) = 0$). Coefficients in level $l$ are normalized by a factor of $\sqrt{2^l}$: this has the effect of making the transform into an *orthonormal basis* [23], so that the sum of squares of coefficients equals the sum of squares of the original data values.

Any data value $g_i$ can be reconstructed as a function of the coefficients which are proper ancestors of the corresponding node in the coefficient tree: the reconstructed value can be found by summing appropriately scaled multiples of these $\log N + 1$ coefficients alone. The *support* of a coefficient $c_i$ is defined as the interval of data values that $c_i$ is used to reconstruct; it is a dyadic interval of size $2^{\log n - l}$ for a coefficient at *resolution level $l$* (see Fig. 1).

Given a limited amount of space for maintaining a *wavelet synopsis* $\mathcal{W}$, a *thresholding procedure* retains a certain number $B \ll n$ of the Haar coefficients as a highly-compressed approximate representation of the original data (the remaining coefficients are implicitly set to 0). Similar to histogram construction, the aim is to determine the "best" subset of $B$ coefficients to retain, so that some overall error measure in the approximation is minimized. By the orthonormality of the normalized Haar basis, greedily picking the $B$ largest coefficients (based on *absolute normalized value*) is provably optimal for the SSE error metric [23]. Recent work proposes schemes for optimal and approximate thresholding under different error metrics. The key idea behind these schemes is to formulate a dynamic program over the coefficient-tree structure that tabulates the optimal solution for a subtree rooted at node $c_j$ *given the contribution from the choices made at the proper-ancestor nodes of $c_j$ in the coefficient tree*. This can handle a broad, natural class of *distributive error metrics* (including, for instance, all the error measures discussed above, as well as weighted $L_p$-norm error for arbitrary $p$) [11, 12].

There are two distinct versions of the thresholding problem for non-SSE error metrics. In the *restricted* version the thresholding algorithm is forced to select values for the synopsis from the standard Haar coeffi-cient values (computed as discussed above). This restriction can lead to sub-optimal synopses for non-SSE error [12]. In the more general, *unrestricted* version of the problem, retained coefficient values are chosen to optimize the target error metric [12]. Let $\text{OPTW}[j, b, v]$ denote the optimal error contribution across all frequencies $g_i$ in the support (i.e., subtree) of coefficient $c_j$ assuming a total space budget of $b$ coefficients retained in $c_j$'s subtree; and, a (partial) reconstructed value of $v$ based on the choices made at proper ancestors of $c_j$. Then, based on the Haar DWT reconstruction process, we can compute $\text{OPTW}[j, b, v]$ as the minimum of two alternative error values at $c_j$:

*(1) Optimal error when retaining the best value for $c_j$,* found by minimizing over all possible values $v_j$ for $c_j$ and allotments of the remaining budget across the left and right child of $c_j$, i.e.

$$\text{OPTW}_r[j, b, v] = \min_{v_j, 0 \leq b' \leq b-1} \{h(\text{OPTW}[2j, b', v + v_j], \text{OPTW}[2j+1, b-b'-1, v-v_j])\}.$$

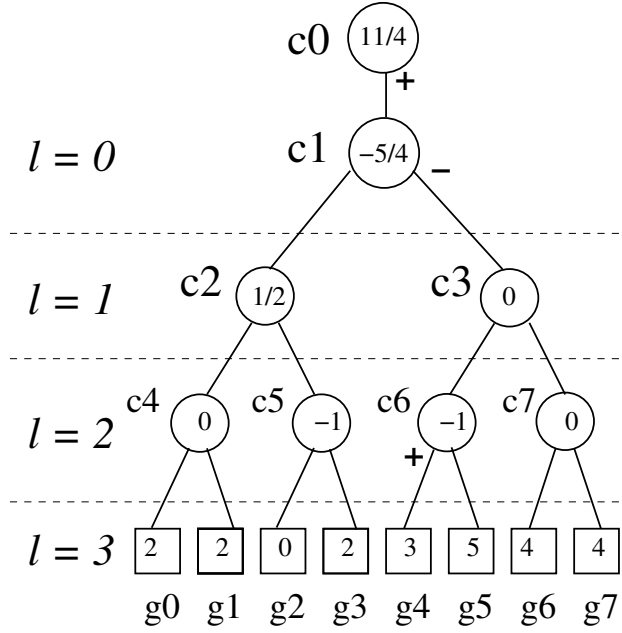

Figure 1: Error-tree structure for data distribution array $A = [2, 2, 0, 2, 3, 5, 4, 4]$, $n = 8$. ($l$ = coefficient resolution levels.)

*(2) Optimal error when not retaining $c_j$,* computed similarly:

$$\text{OPTW}_{nr}[j, b, v] = \min_{v_j, 0 \leq b' \leq b} \{h(\text{OPTW}[2j, b', v], \text{OPTW}[2j+1, b-b', v])\}$$

where $h()$ stands for summation ($\max\{\}$) for cumulative (resp., maximum) error-metric objectives. In the *restricted* problem, the minimization over $v_j$ is eliminated (since the value for $c_j$ is fixed), and the values for the "incoming" contribution $v$ can be computed by stepping through *all possible subsets of ancestors* for $c_j$ — since the depth of the tree is $O(\log n)$, this implies an $O(n^2)$ thresholding algorithm [11]. In the *unrestricted* case, Guha and Harb propose efficient approximation schemes that employ techniques for bounding and approximating the range of possible $v$ values [12].

## 2.3 Probabilistic Data Reduction Problem

The key difference when moving to the probabilistic data setting is that data-distribution frequencies $g_i$ (and, similarly, the corresponding Haar DWT coefficients) are now *random variables* that are instantiated for each individual possible world. We use $g_i(W)$ ($c_i(W)$) to denote the (instantiated) frequency of the $i^{th}$ item (resp., value of the $i^{th}$ Haar coefficient) in possible world $W$ — we sometimes omit the explicit dependence on $W$ for the sake of conciseness. This also implies that the error of a given data synopsis is now a random variable over the collection of possible worlds. Thus, our goal naturally becomes that of constructing a data synopsis $\mathcal{S} \in \{\mathcal{H}, \mathcal{W}\}$ that optimizes *an expected measure of the target error objective over possible worlds*. More formally, let $\mathrm{err}(g_i, \hat{g}_i)$ denote the error of approximating $g_i$ by $\hat{g}_i$ (e.g., squared error or absolute relative error for item $i$); then, our problem can be formulated as follows.

**[Synopsis Construction for Probabilistic Data]** Given a collection of probabilistic attribute values, a synopsis space budget $B$, and a target (cumulative or maximum) error metric, determine a size-$B$ synopsis that minimizes either (1) the expected cumulative error over all possible worlds, i.e., $\mathsf{E}_{\mathcal{W}}[\sum_i \mathrm{err}(g_i, \hat{g}_i)]$ (in the case of a cumulative error objective); or, (2) the maximum value of the per-item expected error over all possible worlds, i.e., $\max_i\{\mathsf{E}_{\mathcal{W}}[\mathrm{err}(g_i, \hat{g}_i)]\}$ (for a maximum error objective). ∎

A natural first attempt to solve the probabilistic data reduction problem is to look to prior work, and ask whether techniques based on sampling, or building a weighted deterministic data set, will suffice. More precisely, one could imagine sampling a possible world $W$ with probability $\Pr[W]$ and building the optimal synopsis for $W$; or for each item $i$ finding $\mathsf{E}_{\mathcal{W}}[g_i]$, and building the synopsis of the "expected" data. Our subsequent analysis shows that such attempts are insufficient. We give precise formulations of the optimal solution to the problem under a variety of error metrics, and one can verify by inspection that they do not in general correspond to any of these simple solutions. Further, we compare the optimal solution to these solutions in our experimental evaluation, and observe that the quality of the solution found is substantially poorer. Observe that this stands in contrast to prior work on estimating functions such as expected number of distinct items [5, 21], which analyzes the number of samples needed to give an accurate estimate. This is because our synopses are not scalar values, and so it is not meaningful to sample many possible worlds, build the synopsis on each, and then find the "average" of these synopses.

## 3 Histograms on Uncertain Data

We primarily consider producing optimal histograms for probabilistic data under cumulative error objectives. These minimize the expected cost of the histogram over all possible worlds. Our techniques are based on applying the dynamic programming approach, and therefore most of our effort is in showing how to to compute the optimal $\hat{b}$ for a bucket $b$ under a given error objective, and also to compute the corresponding value of $\mathsf{E}_{\mathcal{W}}[\mathrm{BERR}(b, \hat{b})]$. Here, we observe that the principle of optimality still holds even under uncertain data: since the expectation of the sum of costs of each bucket is equal to the sum of the expectations, removing the final bucket should leave an optimal $B-1$ bucket histogram over the prefix of the domain. Hence, we will be able to invoke equation (2), and find a solution which requires evaluating $O(Bn^2)$ possibilities.

### 3.1 Sum-squared error

The sum-squared error measure SSE is the sum of the squared differences between the values within a bucket $b_k$ and the representative value of the bucket, $\hat{b}_k$. For a fixed possible world $W$, the optimal value for $\hat{b}_i$ is the mean of the frequencies $g_i$ in the bucket. and the measure reduces to a multiple of the sample variance of the values within the bucket. This holds true even under probabilistic data, as we show below. For deterministic data, it is straightforward to quickly compute the (sample) variance within a given bucket, and therefore use dynamic programming to select the optimum bucketing [19]. In order to use the same approach over uncertain data, which specifies exponentially many possible worlds, we need to be able efficiently compute the variance within a given bucket $b$ specified by start point $s$ and end point $e$.

We first define some necessary quantities. Given the (fixed) discrete frequency distribution implied by $W$, and bucket $b$ of span $n_b$, the sample variance of $W$, $\sigma_b^2(W)$ is defined in terms of $g_i(W)$, the frequency of item $i$ in world $W$, as

$$\sigma_b^2(W) = \sum_{i=s}^{e} \frac{(g_i - \bar{b})^2}{n_b} = \left(\sum_{i=s}^{e} \frac{g_i^2}{n_b}\right) - \left(\sum_{i=s}^{e} \frac{g_i}{n_b}\right)^2 \qquad (3)$$

Given a distribution of possible worlds where $\Pr[W]$ is the probability of possible world $W$, the variance is

$$\mathsf{Var}_{\mathcal{W}}(b) = \mathsf{E}_{\mathcal{W}}[\sigma_b^2] = \sum_{W \in \mathcal{W}} \sigma_b^2(W) \Pr[W] \qquad (4)$$

This follows by substituting the variance function into equation (1).

**Fact 1.** *Under the sum-squared error measure, the cost is minimized by setting $\hat{b} = \mathsf{E}_{\mathcal{W}}[\sum_{i=s}^{e} g_i] = \bar{b}$.*

*Proof.* The cost of the bucket is

$$\mathsf{SSE}(b, \hat{b}) = \mathsf{E}_{\mathcal{W}}[\sum_{i=s}^{e}(g_i - \hat{b})^2] = \mathsf{E}_{\mathcal{W}}[\sum_{i=s}^{e}(g_i - \bar{b} + \hat{b} - \bar{b})^2]$$

$$= \sum_{i=s}^{e} \mathsf{E}_{\mathcal{W}}[(g_i - \bar{b})^2] + \mathsf{E}_{\mathcal{W}}[2(\hat{b} - \bar{b})(g_i - \bar{b}) + (\hat{b} - \bar{b})^2]$$

$$= n_b(\mathsf{Var}_{\mathcal{W}}(b) + 2(\hat{b} - \bar{b})(\mathsf{E}_{\mathcal{W}}[\sum_{i=s}^{e} g_i] - \bar{b}) + (\hat{b} - \bar{b})^2)$$

$$= n_b(\mathsf{Var}_{\mathcal{W}}(b) + (\hat{b} - \bar{b})^2)$$

Since the second term is always positive, it is minimized by setting $\hat{b} = \bar{b}$, yielding the cost as simply $n_b \mathsf{Var}_{\mathcal{W}}(b)$. □

Using equations (3) and (4), we can write $\mathsf{SSE}(b, \bar{b}) = n_b \mathsf{Var}_{\mathcal{W}}(b)$ as the combination of two terms:

$$\mathsf{SSE}(b, \bar{b}) = \sum_{W \in \mathcal{W}} \Pr[W] \sum_{i=s}^{e} g_i(W)^2 - \sum_{W \in \mathcal{W}} \frac{\Pr[W]}{n_b}(\sum_{i=s}^{e} g_i(W))^2$$

$$= \sum_{i=s}^{e} \mathsf{E}_{\mathcal{W}}[g_i^2] - \frac{1}{n_b}\mathsf{E}_{\mathcal{W}}[(\sum_{i=s}^{e} g_i)^2] \qquad (5)$$

The first term is the expectation (over possible worlds) of the sum of squares of frequencies of each item in the bucket. The second term can be interpreted as the expected square of the weight of the bucket, scaled by the span of the bucket. We show how to compute each term efficiently in our different models.

**Value pdf model.** In the value pdf model, we have a distribution for each item $i$ over frequency values $v_j \in \mathcal{V}$ giving $\Pr[g_i = v_j]$. By independence of the values, we have immediately

$$\sum_{i=s}^{e} \mathsf{E}_\mathcal{W}[g_i^2] = \sum_{i=s}^{e} \sum_{v_j \in \mathcal{V}} \Pr[g_i = v_j] v_j^2$$

Since for any random variable $X$ we have $\mathsf{E}[X^2] = \mathsf{Var}[X] + \mathsf{E}[X]^2$, we can use linearity of expectation and summation of variance to find the second term in equation (5) as

$$\mathsf{E}_\mathcal{W}[(\sum_{i=s}^{e} g_i)^2] = \mathsf{E}_\mathcal{W}[\sum_{i=s}^{e} g_i]^2 + \mathsf{Var}_\mathcal{W}[\sum_{i=s}^{e} g_i] \tag{6}$$

$$= (\sum_{i=s}^{e} \sum_{v_j \in \mathcal{V}} \Pr[g_i = v_j] v_j)^2 + \sum_{i=s}^{e} \mathsf{Var}_\mathcal{W}[g_i]$$

where, in turn

$$\mathsf{Var}_\mathcal{W}[g_i] = \sum_{v_j \in \mathcal{V}} \Pr[g_i = v_j] v_j^2 - (\sum_{v_j \in \mathcal{V}} \Pr[g_i = v_j] v_j)^2$$

**Tuple pdf model.** In the tuple pdf case, things appear more involved, because of the interactions between items in the same tuple. As shown by equation (5), we need to compute $\mathsf{E}_\mathcal{W}[g_i^2]$ and $\mathsf{E}_\mathcal{W}[(\sum_i g_i)^2]$. Let the set of tuples in the input be $\mathcal{T} = \{t_j\}$, so that each tuple has an associated pdf giving $\Pr[t_j = i]$, from which can be derived $\Pr[a \leq t_j \leq b]$, the probability that the $i$th tuple in the input is an item between $a$ and $b$ in the input domain.

$$\mathsf{E}_\mathcal{W}[g_i^2] = \mathsf{Var}_\mathcal{W}[g_i] + (\mathsf{E}_\mathcal{W}[g_i])^2$$
$$= \sum_{t_j \in \mathcal{T}} \Pr[t_j = i](1 - \Pr[t_j = i]) + (\sum_{t_j \in \mathcal{T}} \Pr[t_j = i])^2$$

Here, we rely on the fact that variance of each $g_i$ is the sum of the variances arising from each tuple in the input. Observe that although there are dependencies between particular items, these do not affect the computation of the expectations for individual items, which can be then be summed to find the overall answer. For the second term in (5), we can use the expression (6), and we already have an expression for $(\mathsf{E}_\mathcal{W}[\sum_i g_i])^2$. But we cannot simply write $\mathsf{Var}_\mathcal{W}[\sum_{i=s}^{e} g_i]$ as the sum of variances, since these are no longer independent variables. So instead, we treat all items in the same bucket together as a single item and compute its expected square by iterating over all tuples in the input $\mathcal{T}$:

$$\mathsf{Var}_\mathcal{W}[\sum_{i=s}^{e} g_i] = \sum_{t_j \in \mathcal{T}} \Pr[s \leq t_j \leq e](1 - \Pr[s \leq t_j \leq e])$$

**Example.** Consider the input $\langle(1, \frac{1}{2}), (2, \frac{1}{3})\rangle, \langle(2, \frac{1}{4}), (3, \frac{1}{2})\rangle$ in the tuple pdf model from Example 1. We compute

$$\sum_{i=1}^{3} \mathsf{E}_\mathcal{W}[g_i^2] = \frac{1}{4} + \frac{2}{9} + \frac{3}{16} + \frac{1}{4} + \frac{1}{2}^2 + \frac{7}{12}^2 + \frac{1}{2}^2 = \frac{252}{144}.$$

Similarly, $\mathsf{E}_\mathcal{W}[(\sum_{i=1}^{3} g_i)^2] = \frac{5}{36} + \frac{3}{16} + (\frac{5}{6} + \frac{3}{4})^2 = \frac{20+27+361}{144} = \frac{136}{48}$. Combining these, we find the variance of the bucket $1 \ldots 3$ as $\frac{252}{144} - \frac{1}{3} \frac{136}{48} = \frac{29}{36}$. The same value can be obtained by computing the expected sample variance over all possible worlds shown in Example 1. □

**Efficient Computation.** The above development shows how to find the cost of a specified bucket. Computing the minimum cost histogram requires comparing the cost of many different choices of buckets. As in the deterministic case, since the cost is the sum of the costs of the buckets, the dynamic programming solution can find the optimal cost. This computes the cost of the optimal $j$ bucket solution up to position $\ell$ combined with the cost of the optimal $k - j$ bucket solution over positions $\ell + 1$ to $n$. This means finding the cost of $O(n^2)$ buckets. By analyzing the form of the above expressions for the cost of a bucket, we can precompute enough information to allow the cost of any specified bucket to be found in time $O(1)$.

For the tuple pdf model (the value pdf model is similar but simpler) we precompute arrays of length $n$ containing the following information:

$$A[e] = \sum_{i=1}^{e} \Big( \sum_{t_j \in \mathcal{T}} \Pr[t_j = i](1 - \Pr[t_j = i]) + \big( \sum_{t_j \in \mathcal{T}} \Pr[t_j = i] \big)^2 \Big)$$

$$B[e] = \sum_{t_j \in \mathcal{T}} \Pr[t_j \leq e] \qquad C[e] = \sum_{t_j \in \mathcal{T}} (\Pr[t_j \leq e])^2$$

for $1 \leq e \leq s$, and set $A[0] = B[0] = C[0] = 0$. Then the cost $\text{SSE}((s, e), \bar{b})$ is (after some symbolic manipulation) given by

$$A[e] - A[s-1] - \tfrac{(B[e]-B[s-1])(B[e]+B[s-1]+1)-(C[e]-C[s-1])}{(e-s+1)}$$

With the input in sorted order, these three arrays can be computed with a linear pass over the input. We therefore conclude,

**Theorem 1.** *Optimal sum-squared error histograms can be computed over probabilistic data presented in the value pdf or tuple pdf models in time $O(m + Bn^2)$.*

## 3.2 Sum-Squared-Relative-Error

The sum of squares of relative errors measure, SSRE, over deterministic data computes the difference between $\hat{b}$, the representative value for the bucket, and each value within the bucket, and reports the square of this difference as a ratio to the square of the corresponding value. Typically, an additional 'sanity' parameter $c$ is used to limit the value of this quantity in case some values in the bucket are very small. With probabilistic data, we are interested in the expected value of this quantity over all possible worlds. So, given a bucket $b$, the cost is

$$\text{SSRE}(b, \hat{b}) = \mathsf{E}_{\mathcal{W}}[\sum_{i=s}^{e} \frac{(g_i - \hat{b})^2}{\max(c^2, g_i^2)}]$$

By linearity of expectation, the cost for a bucket given $\hat{b}$ can be computed by evaluating at all values of $g_i$ which have a non-zero probability (i.e. at all $v \in \mathcal{V}$).

**Value pdf model.** We can write the sum of squared relative error cost in terms of the probability that, over all possible worlds, the $i$th item has frequency $v_j \in \mathcal{V}$. Then

$$\text{SSRE}(b, \hat{b}) = \sum_{i=s}^{e} \sum_{v_j \in \mathcal{V}} \Pr[g_i = v_j] \frac{(v_j - \hat{b})^2}{\max(c^2, v_j^2)} \qquad (7)$$

We can rewrite this using the function $w(x) = 1/\max(c^2, x^2)$, which is a fixed value once $x$ is specified. Now our cost is

$$\text{SSRE}(b, \hat{b}) = \sum_{i=s}^{e} \sum_{v_j \in \mathcal{V}} \big( \Pr[g_i = v_j] w(v_j) v_j^2 \\ - 2 \Pr[g_i = v_j] w(v_j) v_j \hat{b} + \Pr[g_i = v_j] w(v_j) \hat{b}^2 \big)$$

which is a quadratic in $\hat{b}$. Simple calculus demonstrates that the optimal value of $\hat{b}$ to minimize this cost is

$$\hat{b} = \frac{\sum_{i=s}^{e} \sum_{v_j \in \mathcal{V}} \Pr[g_i = v_j] v_j w(v_j)}{\sum_{i=s}^{e} \sum_{v_j \in \mathcal{V}} \Pr[g_i = v_j] w(v_j)}.$$

Substituting this value of $\hat{b}$ gives $\text{SSRE}(b, \hat{b}) =$

$$\sum_{i=s}^{e} \sum_{v_j \in \mathcal{V}} \Pr[g_i = v_j] v_j^2 w(v_j) - \frac{(\sum_{i=s}^{e} \sum_{v_j \in V} \Pr[g_i = v_j] v_j w(v_j))^2}{\sum_{i=s}^{e} \sum_{v_j \in V} \Pr[g_i = v_j] w(v_j)}$$

In order to compute the cost efficiently, we can compute and store the following arrays:

$$\begin{aligned} X[e] &= \sum_{i=1}^{e} \sum_{v_j \in V} \Pr[g_i = v_j] v_j^2 w(v_j) \\ Y[e] &= \sum_{i=1}^{e} \sum_{v_j \in V} \Pr[g_i = v_j] v_j w(v_j) \\ Z[e] &= \sum_{i=1}^{e} \sum_{v_j \in V} \Pr[g_i = v_j] w(v_j) \end{aligned}$$

From these values, the cost for any given bucket specified by $s$ and $e$ can be computed in constant time as

$$\min_{\hat{b}} \text{SSE}((s, e), \hat{b}) = X[e] - X[s-1] - \frac{(Y[e] - Y[s-1])^2}{Z[e] - Z[s-1]}$$

The standard dynamic programming then finds the optimal set of buckets.

**Tuple pdf model.** Observe that for this cost measure, the cost for the bucket $b$ given by equation (7) is the sum of costs obtained by each item in the bucket. We can focus solely on the contribution to the cost made by a single item $i$, and observe that equation (7) depends only on the (induced) distribution giving $\Pr[g_i = v_j]$: there is no dependency on any other item. As a consequence, we can simply compute the induced value pdf (Section 2.1) for each item independently, and apply the above analysis.

**Theorem 2.** *Optimal sum squared relative error histograms be can computed over probabilistic data presented in the value pdf model in time $O(m + Bn^2)$ and $O(m|\mathcal{V}| + Bn^2)$ in the tuple pdf model.*

### 3.3 Sum-Absolute-Error

As before, $\mathcal{V}$ is the set of possible values taken on by the $g_i$s, indexed so that $v_1 \leq v_2 \leq \ldots \leq v_{|\mathcal{V}|}$. Given some $\hat{b}$, let $j'$ satisfy $v_{j'} \leq \hat{b} < v_{j'+1}$ (we can insert 'dummy' values of $v_0 = 0$ and $v_{|\mathcal{V}|+1} = \infty$ if $\hat{b}$ falls outside of $v_1 \ldots v_{|\mathcal{V}|}$). The sum of absolute errors is given by

$$\text{SAE}(b, \hat{b}) = \sum_{i=s}^{e} \sum_{v_j \in \mathcal{V}} \Pr[g_i = v_j] |\hat{b} - v_j|$$

$$\sum_{i=s}^{e} (\hat{b} - v_{j'}) \Pr[g_i \leq v_{j'}] + (v_{j'+1} - \hat{b}) \Pr[g_i \geq v_{j'+1}]$$

$$+ \sum_{v_j \in \mathcal{V}} \begin{cases} \Pr[g_i \leq v_j](v_{j+1} - v_j) & \text{if } v_j < v_{j'} \\ \Pr[g_i > v_j](v_{j+1} - v_j) & \text{if } v_j \geq v_{j'} \end{cases}$$

The contribution of the first two terms can be written as

$$(v_{j'+1} - v_{j'}) \Pr[g_i \leq v_{j'}] + (\hat{b} - v_{j'+1})(\Pr[g_i \leq v_{j'}] - \Pr[g_i \geq v_{j'+1}])$$

This gives a quantity that is independent of $\hat{b}$ added to one that depends linearly on $\Pr[g_i \leq v_{j'}] - \Pr[g_i > v_{j'+1}]$, which we define as $\Delta_{j'}$. So if $\Delta_{j'} > 0$, we can reduce the cost by making $\hat{b}$ closer to $v_{j'+1}$; if $\Delta_{j'} < 0$, we can reduce the cost by making $\hat{b}$ closer to $v_{j'}$. Therefore, the optimal value of $\hat{b}$ occurs when we make it equal to some $v_j$ value (since, when $\Delta_{j'} = 0$, we have the same result when setting $\hat{b}$ to either $v_{j'}$ or $v_{j'+1}$, or anywhere in between). So we assume that $\hat{b} = v_{j'}$ for some $v_{j'} \in \mathcal{V}$ and can state

$$\text{SAE}(b, \hat{b}) = \sum_{i=s}^{e} \sum_{v_j \in \mathcal{V}} \begin{cases} \Pr[g_i \leq v_j](v_{j+1} - v_j) & \text{if } \hat{b} > v_j \\ \Pr[g_i > v_j](v_{j+1} - v_j) & \text{if } \hat{b} \leq v_j \end{cases}$$

Define $P_{j,s,e} = \sum_{i=s}^{e} \Pr[g_i \leq v_j]$ and $P^*_{j,s,e} = \sum_{i=s}^{e} \Pr[g_i > v_j]$. Observe that $P_{j,s,e}$ is monotone increasing in $j$ while $P^*_{j,s,e}$ is monotone decreasing in $j$. So we have

$$\text{SAE}(b, \hat{b}) = \sum_{v_j < \hat{b}} P_{j,s,e}(v_{j+1} - v_j) + \sum_{v_j \geq \hat{b}} (P^*_{j,s,e})(v_{j+1} - v_j) \tag{8}$$

Now observe that we have a contribution of $(v_{j+1} - v_j)$ for all $j$ values, multiplied by either $P[j, s, e]$ or $P^*[j, s, e]$. Consider the effect on SAE if we step $\hat{b}$ through values $v_1, v_2 \ldots v_{|\mathcal{V}|}$. We have

$$\text{SAE}(b, v_{\ell+1}) - \text{SAE}(b, v_\ell) = (P_{\ell,s,e} - P^*_{\ell+1,s,e})(v_{\ell+1} - v_\ell)$$

Because $P_{j,s,e}$ is monotone increasing in $j$, and $P^*_{j,s,e}$ is monotone decreasing in $j$, the quantity $P_{\ell,s,e} - P^*_{\ell+1,s,e}$ is monotone increasing in $\ell$. Thus $\text{SAE}(b, \hat{b})$ can have a single minimum value as $\hat{b}$ is varied, and is increasing in both directions away from this value. This minimum does not depend on the $v_j$ values themselves; instead, it occurs (approximately) when $P_{j,s,e} \approx P^*_{j,s,e} \approx n_b/2$, where $n_b = (e - s + 1)$ as before. It therefore suffices to find the $v'_j$ value defined by

$$v'_j = \arg\min_{v_\ell \in \mathcal{V}} \sum_{v_j < v_\ell \in \mathcal{V}} P_{j,s,e} + \sum_{v_j > v_\ell \in \mathcal{V}} P^*_{j,s,e}$$

and then set $\hat{b} = v_{j'}$ to obtain the optimal SAE cost.

**Tuple and Value pdf models.** In the value pdf case, it is straightforward to compute the $P$ and $P^*$ values directly from the input pdfs. For the tuple pdf case, observe that from the form of the expression for SAE, there are no interactions *between* different $g_i$ values: although the input specifies interactions and (anti)correlations between different variables, for computing the error in a bucket we can treat each item independently in turn. We therefore convert to the induced value pdf (at an additional cost of $O(m|\mathcal{V}|)$), and use this in our subsequent computations.

**Efficient Computation.** In order to quickly find the cost of a given bucket $b$, we must find the optimal $\hat{b}$ and the cost of the bucket using $\hat{b}$ as the representative. Our approach is to precompute $\sum_{v_j < \ell} P_{j,1,e}(v_{j+1} - v_j)$ and $\sum_{v_j \geq \ell} P^*_{j,1,e}(v_{j+1} - v_j)$ values for all values of $v_\ell \in \mathcal{V}$ and $e \in [n]$. Now $\text{SAE}(b, \hat{b})$ for any $\hat{b} \in \mathcal{V}$ can be computed (from (8)) as the sum of two differences of precomputed values. The minimum value attainable by any $\hat{b}$ can then be found by a ternary search over the values $\mathcal{V}$, using $O(\log |\mathcal{V}|)$ probes. Finally, the cost for the bucket using this $\hat{b}$ is also found from the same information. The cost is $O(|\mathcal{V}|n)$ preprocessing to build tables of prefix sums, and then $O(\log |\mathcal{V}|)$ to find the optimal cost of a given bucket. Therefore,

**Theorem 3.** *Optimal sum-absolute-error histograms can be computed over probabilistic data presented in the (induced) value pdf model in time $O(n(|\mathcal{V}| + Bn + n \log |\mathcal{V}|))$.*

Note that for all of models of probabilistic data, $|\mathcal{V}| \leq m$ is polynomial in the size of the input, so this is a fully polynomial time algorithm.

### 3.4 Sum-Absolute-Relative-Error

For sum of absolute relative errors, the bucket cost $\text{SARE}(b, \hat{b})$ is

$$\mathsf{E}_\mathcal{W}(\sum_{i=s}^{e} \frac{|g_i - \hat{b}|}{\max(c, g_i)}) = \sum_{i=s}^{e} \sum_{v_j \in \mathcal{V}} \frac{\Pr[g_i = v_j]}{\max(c, v_j)} |v_j - \hat{b}|$$
$$= \sum_{i=s}^{e} \sum_{v_j \in \mathcal{V}} w_{i,j} |v_j - \hat{b}|$$

where we define $w_{i,j} = \frac{\Pr[g_i = v_j]}{\max(c, v_j)}$. But, more generally, the $w_{i,j}$ can be arbitrary non-negative weights. Setting $j'$ so that $v_{j'} \leq \hat{b} \leq v_{j'+1}$, we can write the cost as

$$\sum_{i=s}^{e} \sum_{v_j \in \mathcal{V}} \begin{cases} w_{i,j}(\hat{b} - v_j) & \text{if } v_j < \hat{b} \\ w_{i,j}(v_j - \hat{b}) & \text{if } v_j \geq \hat{b} \end{cases}$$

$$= \sum_{i=s}^{e} \sum_{v_j \in \mathcal{V}} \begin{cases} w_{i,j}(\hat{b} - v_{j'} + \sum_{v_j \leq v_\ell < v_{j'}} v_{\ell+1} - v_\ell) & \text{if } v_j < \hat{b} \\ w_{i,j}(v_{j'} - \hat{b} + \sum_{v_{j'} \leq v_\ell < v_j} v_{\ell+1} - v_\ell) & \text{if } v_j \geq \hat{b} \end{cases} \quad (9)$$

We define $W_{i,j} = \sum_{r=1}^{j} w_{i,r}$ and $W^*_{i,j} = \sum_{r=j+1}^{|\mathcal{V}|} w_{i,r}$, so that rearranging the previous sum, the cost is

$$\text{SARE}(b, \hat{b}) = \sum_{i=s}^{e} W_{i,j'}(\hat{b} - v_{j'}) - W^*_{i,j}(\hat{b} - v_{j'+1})$$

$$+ \sum_{v_j \in \mathcal{V}} \begin{cases} W_{i,j}(v_{j+1} - v_j) & \text{for } v_{j'} > v_j \\ W^*_{i,j}(v_j - v_{j-1}) & \text{for } v_{j'} \leq v_j \end{cases}$$

The same style of argument as in the previous section suffices to show that the optimal choice of $\hat{b}$ is when $\hat{b} = v_{j'}$ for some $j'$. We define $P_{j,s,e} = \sum_{i=s}^{e} W_{i,\ell}$ and $P^*_{j,s,e} = \sum_{i=s}^{e} W^*_{i,\ell}$ so we have

$$\text{SARE}(b, \hat{b}) = \sum_{v_{j'} > v_j \in V} P_{j,s,e}(v_{j+1} - v_j) + \sum_{v_{j'} \leq v_j \in V} P^*_{j,s,e}(v_{j+1} - v_j).$$

Observe that this matches the form of (8). As in Section 3.3, $P_{j,s,e}$ is monotone increasing in $j$, and $P^*_{j,s,e}$ is decreasing in $j$. Therefore, the same argument holds to show that there is a unique minimum value of SARE, and it can be found by a ternary search over the range. Likewise, the form of the cost in equation (9) shows that there are no interactions between different items, and so we can work in the (induced) value pdf model. By building corresponding data structures based on tabulating prefix sums of the new $P$ and $P^*$ functions, we conclude:

**Theorem 4.** *Optimal sum absolute relative error histograms can be computed over probabilistic data presented in the tuple and value pdf models in time $O(n(|\mathcal{V}| + Bn + n \log |\mathcal{V}| \log n))$.*

## 3.5 Approximate Histogram Computation

Our results so far all cost at least $\Omega(Bn^2)$ due to the use of dynamic programming to find the optimal bucket boundaries. As observed in prior work, it is not always profitable to expend so much effort if the resulting histogram is to be used to approximate the original input; clearly, if we tolerate approximation in this way, then we should also be able to tolerate a histogram which achieves close to the optimal cost rather than exactly the optimal. In particular, we should be happy if we can find a histogram whose cost is at most $(1 + \epsilon)$ times the cost of the optimal histogram in time much faster than $\Omega(Bn^2)$.

Here, we can adopt the approach of Guha *et al.* [13, 14]. Instead of considering every possible bucket, we can use properties of the error measure, and consider only a subset of possible buckets, much accelerating the search. We observe that the following conditions hold for all the previously considered error measures: (1) The error of a bucket only depends on the size of the bucket and the distributions of the items falling within it; (2) The overall error is the sum of the errors across all buckets; (3) We can maintain information so that given any bucket $b$ the best representative $\hat{b}$ and corresponding error can be computed efficiently; (4) The error is monotone, so that the error for any interval of items is no less than the error of any contained subinterval; and (5) The total error cost is bounded as a polynomial in the size of the input.

Note that most of our work so far has been in giving analysis and techniques to support point (3) above; the remainder of the points are simple consequences of the definition of the error measure. As a result of these properties, we invoke Theorem 6 of [14], and state

**Theorem 5.** *Given preprocessing as described in previous sections, we find a $(1 + \epsilon)$ approximation to the optimal histogram for SSE, SSRE, SAE and SARE, with $O(\frac{1}{\epsilon}B^2 n \log n)$ bucket cost evaluations using the algorithm described in [14].*

## 3.6 Maximum-Absolute-Error and Maximum-Absolute-Relative-Error

Thus far we have relied on the linearity of expectation and related properties such as summability of variance to simplify the expressions of cost and aid in the analysis. When we consider other error metrics, such as the maximum error and maximum relative error, we cannot immediately use such linearity properties, and so the task becomes more involved. Here, we provide results for maximum absolute error and maximum absolute relative error, MAE and MARE. Over a deterministic input, the maximum error in a bucket $b$ is $\text{MAE}(b, \hat{b}) = \max_{s \leq i \leq e} |g_i - \hat{b}|$, and the maximum relative error is $\text{MARE}(b, \hat{b}) = \max_{s \leq i \leq e} \frac{|g_i - \hat{b}|}{\max(c, g_i)}$. We focus on bounding the maximum value of the per-item expected error[1]. Here, we consider the frequency

---
[1]Note that the alternate formulation, where we seek to minimize the expectation of the maximum error, is also plausible, and worthy of further study.

of each item in the bucket in turn for the expectation, and then take the maximum over these costs. So we can write the costs as

$$\text{MAE}(b, \hat{b}) = \max_{s \leq i \leq e} \sum_{v_j \in \mathcal{V}} \Pr[g_i = v_j] |v_j - \hat{b}|$$

$$\text{MARE}(b, \hat{b}) = \max_{s \leq i \leq e} \sum_{v_j \in \mathcal{V}} \frac{\Pr[g_i = v_j]}{\max(c, v_j)} |v_j - \hat{b}|$$

We can represent these both as $\max_{s \leq i \leq e} \sum_{j=1}^{|\mathcal{V}|} w_{i,j} |v_j - \hat{b}|$, where $w_{i,j}$ are non-negative weights independent of $\hat{b}$. Now observe that we have the maximum over what can be thought of as $n_b$ parallel instances of a sum-absolute relative error (SARE) problem, one for each $i$ value. Following the analysis in Section 3.4, we observe that each function $f_i(b) = \sum_{j=1}^{|\mathcal{V}|} |v_j - \hat{b}|$ has a single minimum value, and is increasing away from its minimum. It follows that the upper envelope of these functions, given by $\max_{s \leq i \leq e} f_i(b)$ also has a single minimum value, and is increasing as we move away from this minimum. So we can perform a ternary search over the values of $v_j$ to find $j'$ such that the optimal $\hat{b}$ lies between $v_{j'}$ and $v_{j'+1}$. Each evaluation for a chosen value of $\hat{b}$ can be completed in time $O(n_b)$: that is, $O(1)$ for each value of $i$, by creating the appropriate prefix sums as discussed in Section 3.4 (it is possible to improve this cost by appropriate pre-computations, but this will not significantly alter the asymptotic cost of the whole operation). The ternary search over the values in $\mathcal{V}$ takes $O(\log |\mathcal{V}|)$ evaluations, giving a total cost of $O(n_b \log |\mathcal{V}|)$.

Knowing that $\hat{b}$ must lie in this range, the cost is of the form

$$\text{MARE}(b, \hat{b}) = \max_{s \leq i \leq e} \alpha_i(\hat{b} - v_{j'}) + \beta_i(v_{j'+1} - \hat{b}) + \gamma_i$$
$$= \max_{s \leq i \leq e} \hat{b}(\alpha_i - \beta_i) + (\gamma_i + \beta_i v_{j'+1} - \alpha_i v_{j'})$$

where the $\alpha_i, \beta_i, \gamma_i$ values are determined solely by $j'$, the $w_{i,j}$'s and the $v_j$s, and are independent of $\hat{b}$. This means we must now minimize the maximum value of a set of univariate linear functions in the range $v_{j'} \leq \hat{b} \leq v_{j'+1}$. A divide-and-conquer approach, based on recursively finding the intersection of convex hulls of subsets of the linear functions yields an $O(n_b \log n_b)$ time algorithm[2]. Combining these, we determine that evaluating the optimal $\hat{b}$ and the corresponding cost for a given bucket takes time $O(n_b \log n_b |\mathcal{V}|)$. We can then apply the dynamic programming solution, since the principle of optimality holds over this error objective. Because of the structure of the cost function, it suffices to move from the tuple pdf model to the induced value pdf, and so we conclude,

**Theorem 6.** *The optimal $B$ bucket histogram under maximum-absolute-error (MAE) and maximum-absolute-relative-error (MARE) over probabilistic data in either the tuple or value pdf models can be found in time $O(n^2(B + n \log n |\mathcal{V}|))$.*

## 4 Wavelets on Probabilistic Data

We first present our results on the core problem of finding the $B$ term optimal wavelet representation under sum-squared error, and then discuss extensions to other error objectives.

---

[2]The same underlying optimization problem arises in a weighted histogram context [15], which gives full details of this approach.

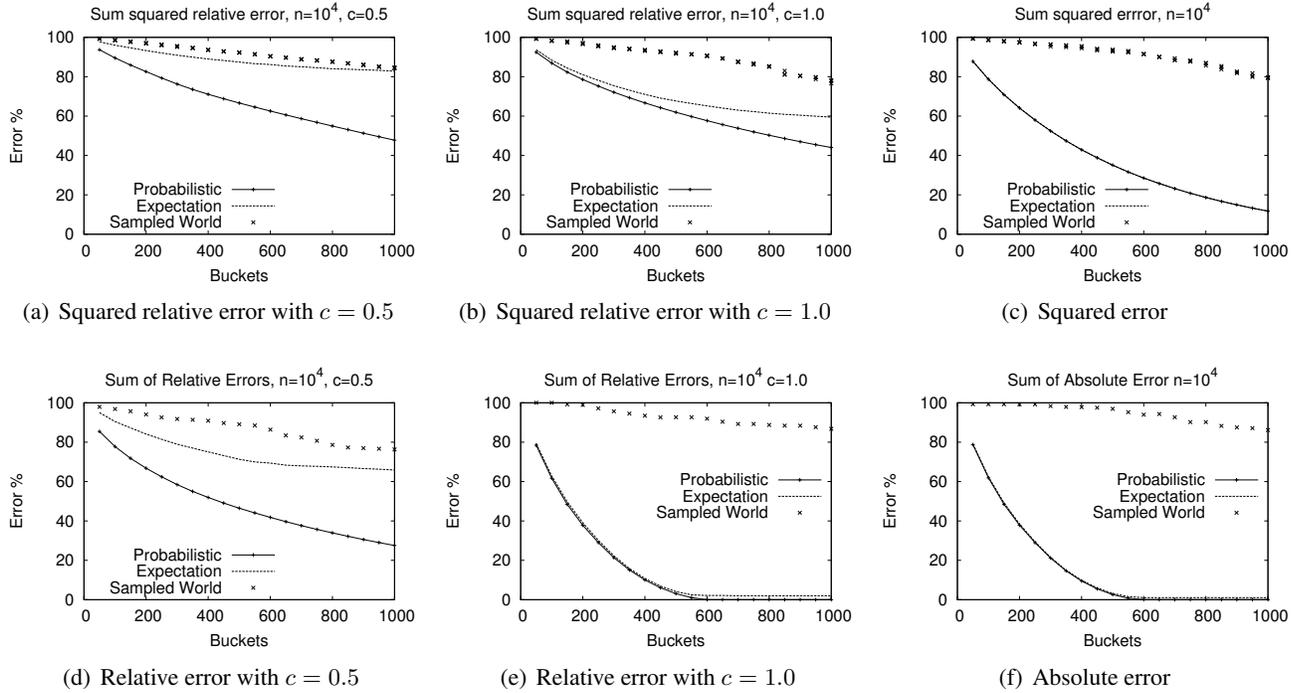

Figure 2: Results on Histogram Computation

## 4.1 SSE-Optimal Wavelet Synopses.

Any input which defines a distribution over original $g_i$ values immediately implies a distribution over Haar wavelet coefficients $c_i$. In particular, we have a possible-worlds distribution over Haar DWT coefficients, with $c_i(W)$ denoting the instantiation of $c_i$ in world $W$ (defined by the $g_i(W)$s). Our goal is to pick a set of $B$ coefficient indices $\mathcal{I}$ and corresponding coefficient values $\hat{c}_i$ for each $i \in \mathcal{I}$ so as to minimize the expected SSE in the data approximation. By Parseval's theorem [23] and the linearity of the wavelet transform, in each possible world, the SSE of the data approximation is simply the SSE in the approximation of the normalized wavelet coefficients. Thus, by linearity of expectation, the expected SSE for the resulting synopsis $\mathcal{S}_w(\mathcal{I})$ is:

$$\mathsf{E}_{\mathcal{W}}[\mathrm{SSE}(\mathcal{S}_w(\mathcal{I}))] = \sum_{i \in \mathcal{I}} \mathsf{E}_{\mathcal{W}}[(c_i - \hat{c}_i)^2] + \sum_{i \notin \mathcal{I}} \mathsf{E}[(c_i)^2].$$

Some observations follow immediately. Suppose we are to include $i$ in our index set $\mathcal{I}$ of selected coefficients. Then the optimal setting of $\hat{c}_i$ is the expected value of the $i^{th}$ (normalized) Haar wavelet coefficient, by the same argument style as Fact 1. That is,

$$\mu_{c_i} = \mathsf{E}_{\mathcal{W}}[c_i] = \sum_{w_j} \Pr[c_i = w_j] \cdot w_j,$$

computed over the set of values taken on by the coefficients, $w_j$. Further, by linearity of expectation and the fact that the Haar wavelet transform can be thought of as a linear operator $H$ applied to the input vector $A$, we have

$$\mu_{c_i} = \mathsf{E}_{\mathcal{W}}[H_i(A)] = H_i(\mathsf{E}_{\mathcal{W}}[A]).$$

In other words, we can find the $\mu_{c_i}$'s by computing the wavelet transform of the expected frequencies,

$E_\mathcal{W}(g_i)$. So the $\mu_{c_i}$s can be computed with linear effort from the input, in either tuple pdf or value pdf form. Based on the above observation, we can rewrite the expected SSE as:

$$E_\mathcal{W}[\text{SSE}(\mathcal{S}_w(\mathcal{I}))] = \sum_{i \in \mathcal{I}} \sigma_{c_i}^2 + \sum_{i \notin \mathcal{I}} E[(c_i)^2],$$

where $\sigma_{c_i}^2 = \text{Var}_\mathcal{W}[c_i]$ is the variance of $c_i$. From the above expression, it is clear that the optimal strategy is to pick the $B$ coefficients giving the largest reduction in the expected SSE (since there are no interactions across coefficients); furthermore, the "benefit" of selecting coefficient $i$ is exactly $E[(c_i)^2] - \sigma_{c_i}^2 = \mu_{c_i}^2$. Thus, the thresholding scheme that optimizes expected SSE is to simply select the $B$ Haar coefficients with the largest (absolute) expected normalized value. (It is interesting to note that this scheme naturally generalizes the conventional deterministic SSE thresholding case (Section 2.2).)

**Theorem 7.** *With $O(n)$ time and space, we can compute the optimal* SSE *wavelet representation of probabilistic data in the tuple and value pdf models.*

### 4.2 Wavelet Synopses for non-SSE Error.

The DP recurrence formulated over the Haar coefficient error tree for non-SSE error metrics in the deterministic case (Section 2.2) extends naturally to the case of probabilistic data. The only change is that we now define $\text{OPTW}[j, b, v]$ to denote the *expected* optimal value for the error metric of interest under the same conditions as the deterministic case. The recursive computation steps remain exactly the same. The interesting point with the coefficient-tree DP recurrence is that almost all of the actual error computation takes place at the *leaf* (i.e., data) nodes of the tree — the DP recurrences simply combine these computed error values appropriately in a bottom-up fashion. For deterministic data, the error at a leaf node $i$ with an incoming value of $v$ from its parents is just the point error metric of interest with $\hat{g}_i = v$; that is, for leaf $i$, we simply compute $\text{OPTW}[i, 0, v] = \text{err}(g_i, v)$ which can be done trivially in $O(1)$ time (note that leaf entries in $\text{OPTW}[]$ are only defined for $b = 0$ since space is never allocated to leaves).

In the case of probabilistic data, such leaf-error computations are a little more complicated since we now need to compute the *expected point-error value*

$$E_\mathcal{W}[\text{err}(g_i, v)] = \sum_W \Pr[W] \cdot \text{err}(g_i(W), v)$$

over all possible worlds $W \in \mathcal{W}$. Fortunately, this computation can still be done in $O(1)$ time assuming some simple precomputed data structures, similar to those we have derived for error objectives in the histogram case. To illustrate the main ideas, consider the case of *absolute relative error metrics*, i.e., $\text{err}(g_i, \hat{g}_i) = w(g_i) \cdot |g_i - \hat{g}_i|$ where $w(g_i) = 1/\max\{c, |g_i|\}$. Then, we can expand the expected error at $g_i$ as follows:

$$\begin{aligned}\text{OPTW}[i, 0, v] &= E_\mathcal{W}[w(g_i) \cdot |g_i - v|] \\ &= \sum_{v_j \in \mathcal{V}} \Pr[g_i = v_j] w(v_j) \cdot \begin{cases} (v - v_j) & \text{if } v > v_j \\ (v_j - v) & \text{if } v \leq v_j \end{cases}\end{aligned}$$

where, as earlier, $\mathcal{V}$ denotes the set of possible values for any frequency random variable $g_i$. In other words, we have an instance of a sum-absolute-relative-error problem, since the form of this optimization matches that in Section 3.3. So by precomputing appropriate arrays of size $O(|\mathcal{V}|)$ for each $i$, we can search for the optimal "split point" $v_{j'} \in \mathcal{V}$ in time $O(\log |\mathcal{V}|)$.

The above precomputation ideas can naturally be extended to other error metrics as well, and allow us to easily carry over the algorithms and results (modulo the small $O(\log |\mathcal{V}|)$ factor above) for the *restricted* case, where all coefficient values are fixed, e.g., to their expected values as required for expected SSE minimization. The following theorem summarizes our discussion.

**Theorem 8.** *Optimal restricted wavelet synopses for non-*SSE *error metrics can be computed over probabilistic data presented in the (induced) value pdf model in time* $O(n(|\mathcal{V}| + n \log |\mathcal{V}|))$.

For the *unrestricted* case, some additional work is needed in order to effectively bound and quantize the range of possible coefficient values to consider in the case of probabilistic data at the leaf nodes of the tree. One option is to consider pessimistic coefficient-range estimates (e.g., based on the minimum/maximum possible frequency values); another option would be to employ some tail bounds on the $g_i$'s (e.g., using Chernoff since tuples can be seen as binomial variables) in order to derive *tighter, high-probability* ranges for coefficient values. We defer a more detailed exploration to the full version of this paper.

## 5 Experiments

We implemented our algorithms in C, and carried out a set of experiments to compare the quality and scalability of our results against those from naively applying methods designed for deterministic data. Experiments were performed on a desktop 2.4GHz machine with 2GB RAM.

**Data Sets.** We experimented using a mixture of real and synthetic data sets. The real dataset came from the MystiQ project[3] which includes approximately $m = 127,000$ tuples describing $27,700$ distinct items. These correspond to links between a movie database and an e-commerce inventory, so the tuples for each item define the distribution of the number of expected matches. This uncertain data provides input in the basic model. the items for various subsets of the relation. Synthetic data was generated using the MayBMS [1] extension to the TPC-H generator [4]. We used the lineitem-partkey relation, where the multiple possibilities for each uncertain item are interpreted as tuples with uniform probability over the set of values in the tuple pdf model.

**Sampled Worlds and Expectation.** We compare our methods to two naive methods of building a synopsis for uncertain data using deterministic techniques discussed in Section 2.3. The first is to simply sample a possible world, and compute the (optimal) synopsis for this deterministic sample. The second is to compute the expected frequency of each item, and build the synopsis of this deterministic input. This can be thought of as equivalent to sampling many possible worlds, combining and scaling the frequencies of these, and building the summary of the result. For consistency, we use the same code to compute the respective synopses over both probabilistic and certain data, since deterministic data can be interpreted as probabilistic data in the value pdf model with probability 1 of attaining a certain frequency.

### 5.1 Histograms on Probabilistic Data

We use our methods described in Section 3 to build the histogram over $n$ items using $B$ buckets, and compute the cost of the histograms under the relevant metric (e.g. the expected sum-relative-error, etc.). Observe that, unlike in the deterministic case, a histogram with $B = n$ buckets does not have zero error: we have to choose a fixed representative $\hat{b}$ for each bucket, so any bucket with some uncertainty will have a contribution to the expected cost. We therefore compute the percentage error of a given histogram as the fraction of the cost difference between the one bucket histogram (largest achievable error) and the $n$ bucket histogram (smallest achievable error).

---

[3] http://www.cs.washington.edu/homes/suciu/project-mystiq.html

[4] www.cs.cornell.edu/database/maybms/

**Quality.** For uniformity, we show results for the MystiQ movie data set; results on synthetic data were similar, and are omitted for space reasons. The quality of the different methods on the same $n = 10,000$ distinct data items is shown in Figure 2. In each case, we measured the cost for using up to $1000$ buckets over the four cumulative error measures: sum-squared-error, sum-squared-relative-error, sum-absolute-error and sum-relative-error. We show two values of the sanity constant for relative error, $c = 0.5$ and $c = 1.0$. Since our results show that the dynamic programming finds the optimum set of buckets, there is no surprise that the cost is always smaller than the two naive methods. Figure 2(a) shows a typical case for relative error: the probabilistic method is appreciably better than using the expected costs, which in turn is somewhat better than building the histogram of a sampled world. We show the results for three independent samples to show that there is fairly little variation in the cost. For the sum-squared-error and sum-absolute-error (Figure 2(c) and 2(f)), while using a sampled world is still poor, the cost of using the expectation is very close to that of our probabilistic method. The reason is that the histogram obtains the most benefit by putting items with similar behavior in the same bucket, and on this data, the expectation is a good indicator of behavioral similarity. This is not always the case, and indeed, Figure 2(f) shows that while our method obtains the smallest possible error with about 600 buckets, using the expectation finds a bucketing with a slightly higher cost. Other values of $c$ tend to vary smoothly between two extremes: increasing $c$ allows the expectation method to get closer to the probabilistic solution (as in Figure 2(e)). This is because as $c$ approaches the maximum achievable frequency of any item, there is no longer any dependency on the frequency, only on $c$, and so the cost function is essentially a scaled version of the squared error or absolute error respectively. Reducing $c$ towards 0 further disadvantages the expectation method, and it has close to 100% error even when a very large number of buckets are provided; meanwhile, the probabilistic method smoothly reduces in error down to zero.

**Scalability.** Figure 3 shows the time cost of our methods. We show the results for sum squared error, although the results are very similar for other metrics, due to a shared code base. We see a strong linear dependency on the number of buckets, $B$, and a close to quadratic dependency on $n$ (since as $n$ doubles, the time cost slightly less than quadruples). This confirms our analysis that shows the cost is dominated by an $O(Bn^2)$ term. The time to apply the naive methods is almost identical, since they both ultimately rely on solving a dynamic program of the same size, which dwarfs any reduced linear preprocessing cost. Therefore, the cost is essentially the same as for deterministic data. The time cost is acceptable, but it suggests that for larger relations it will be advantageous to pursue the faster approximate solutions outlined in Section 3.5.

## 5.2 Wavelets on Probabilistic Data

We implemented our methods for computing wavelets under the sum-squared-error (SSE) objective. Here, the analysis shows that the optimal solution is to compute the wavelet representation of the expected data (since this is equivalent to generating the expected values of the coefficients, due to linearity of the wavelet transform function), and then pick the $B$ largest coefficients. We contrast to the effect of sampling possible worlds and picking the coefficients corresponding to the largest coefficients of the sampled data. We measure the error by computing the sum of the square of the $\mu_{c_i}$s not picked by the method, and expressing this as a percentage of the sum of all such $\mu_{c_i}$s, since our analysis demonstrates that this is the range of possible error (Section 4.1). Figures 4(a) and 4(b) shows the effect of varying the number of coefficients $B$ on the real and synthetic data sets: while increasing the number of coefficients improves the cost in the sampled case, it is much more expensive than the optimal solution. Both approaches take the same amount of time, since they rely on computing a standard Haar wavelet transform of certain deterministic data: it takes linear time to produce the expected values and then compute the coefficients; these are sorted to find the $B$ largest. This took much less than a second on our experimental set up.

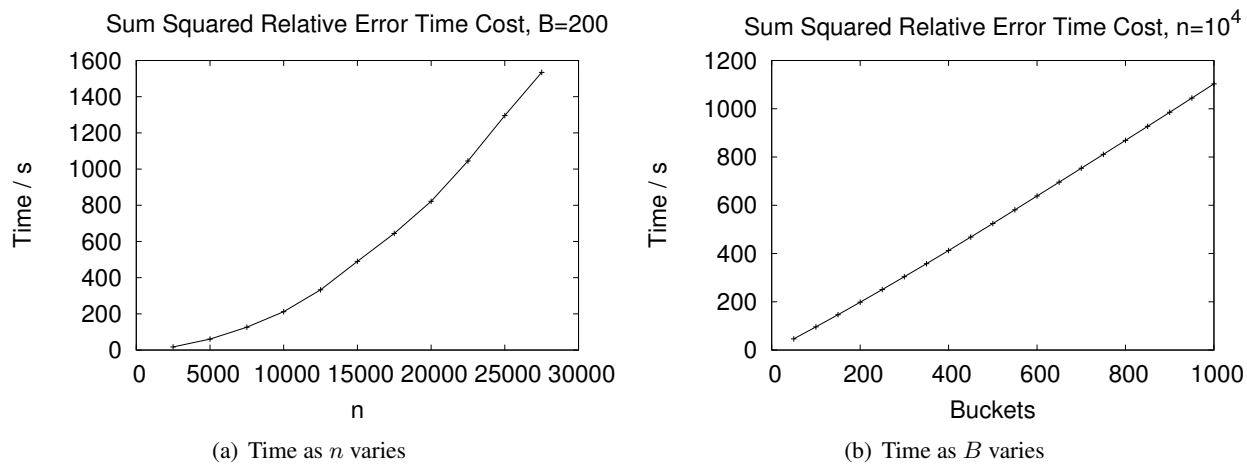

(a) Time as $n$ varies   (b) Time as $B$ varies

Figure 3: Histogram Timing Costs

## 6 Concluding Remarks

We have introduced the probabilistic data reduction problem, and given results for a variety of cumulative and maximum error objectives, for both histograms and wavelets. Empirically, we see that the optimal synopses can accurately summarize the data and are significantly better than simple heuristics. It remains to better understand how to approximately represent and process probabilistic data. So far, we have focused on the foundational one-dimensional problem, but is also important to study multi-dimensional generalizations. The error objective formulations we have analyzed implicitly assume uniform workloads for point queries, and so it remains to address the case when in addition to a distribution over the input data, there is also a distribution over the queries to be answered.

**Acknowledgements.** We thank Sudipto Guha for some useful suggestions; Dan Suciu for providing data from the MystiQ project, and Dan Olteanu for generating data with the MayBMS system.

## References

[1] L. Antova, T. Jansen, C. Koch, and D. Olteanu. Fast and simple relational processing of uncertain data. In *IEEE International Conference on Data Engineering*, 2008.

[2] O. Benjelloun, A. D. Sarma, C. Hayworth, and J. Widomn. An introduction to ULDBs and the Trio system. *IEEE Data Engineering Bulletin*, 29(1):5–16, Mar. 2006.

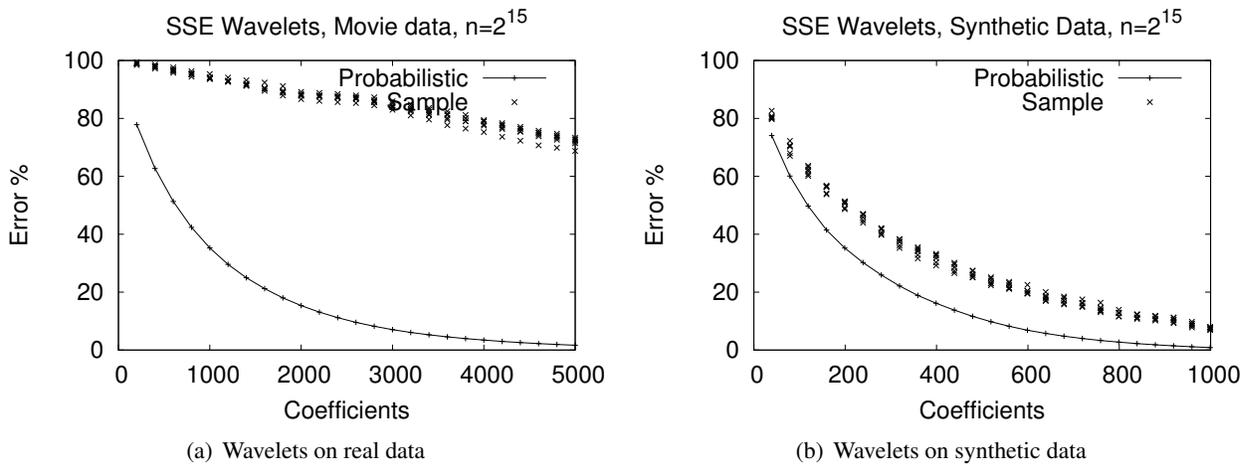

Figure 4: Results on Wavelet Computation

[3] J. Boulos, N. Dalvi, B. Mandhani, S. Mathur, C. Re, and D. Suciu. Mystiq: A system for finding more answers by using probabilities. In *ACM SIGMOD International Conference on Management of Data*, 2005.

[4] K. Chakrabarti, M. Garofalakis, R. Rastogi, and K. Shim. Approximate query processing using wavelets. In *International Conference on Very Large Data Bases*, 2000.

[5] G. Cormode and M. Garofalakis. Sketching probabilistic data streams. In *ACM SIGMOD International Conference on Management of Data*, 2007.

[6] G. Cormode and A. McGregor. Approximation algorithms for clustering uncertain data. In *ACM Principles of Database Systems*, 2008.

[7] N. Dalvi and D. Suciu. Efficient query evaluation on probabilistic databases. In *International Conference on Very Large Data Bases*, 2004.

[8] N. N. Dalvi and D. Suciu. Management of probabilistic data: foundations and challenges. In *ACM Principles of Database Systems*, 2007.

[9] M. Garofalakis and P. B. Gibbons. Approximate query processing: Taming the terabytes. In *International Conference on Very Large Data Bases*, 2001.

[10] M. Garofalakis and P. B. Gibbons. Probabilistic wavelet synopses. *ACM Transactions on Database Systems*, 29(1), Mar. 2004.

[11] M. Garofalakis and A. Kumar. Wavelet synopses for general error metrics. *ACM Transactions on Database Systems*, 30(4), Dec. 2005.

[12] S. Guha and B. Harb. Wavelet synopsis for data streams: minimizing non-euclidean error. In *ACM SIGKDD*, 2005.

[13] S. Guha, N. Koudas, and K. Shim. Data streams and histograms. In *ACM Symposium on Theory of Computing*, pages 471–475, 2001.

[14] S. Guha, N. Koudas, and K. Shim. Approximation and streaming algorithms for histogram construction problems. *ACM Transactions on Database Systems*, 31(1):396–438, 2006.

[15] S. Guha and K. Shim. A note on linear time algorithms for maximum error histograms. *IEEE Transactions on Knowledge and Data Engineering*, 19(7):993–997, 2007.

[16] S. Guha, K. Shim, and J. Woo. REHIST: Relative error histogram construction algorithms. In *International Conference on Very Large Data Bases*, 2004.

[17] Y. E. Ioannidis. The history of histograms (abridged). In *International Conference on Very Large Data Bases*, 2003.

[18] Y. E. Ioannidis and V. Poosala. Balancing histogram optimality and practicality for query result size estimation. In *ACM SIGMOD International Conference on Management of Data*, pages 233–244, 1995.

[19] H. V. Jagadish, N. Koudas, S. Muthukrishnan, V. Poosala, K. Sevcik, and T. Suel. Optimal histograms with quality guarantees. In *International Conference on Very Large Data Bases*, 1998.

[20] T. S. Jayram, S. Kale, and E. Vee. Efficient aggregation algorithms for probabilistic data. In *ACM-SIAM Symposium on Discrete Algorithms*, 2007.

[21] T. S. Jayram, A. McGregor, S. Muthukrishnan, and E. Vee. Estimating statistical aggregates on probabilistic data streams. In *ACM Principles of Database Systems*, 2007.

[22] N. Khoussainova, M. Balazinska, and D. Suciu. Towards correcting input data errors probabilistically using integrity constraints. In *ACM Workshop on Data Engineering for Wireless and Mobile Access (MobiDE)*, 2006.

[23] E. J. Stollnitz, T. D. DeRose, and D. H. Salesin. *"Wavelets for Computer Graphics – Theory and Applications"*. Morgan Kaufmann Publishers, San Francisco, CA, 1996.

[24] D. Suciu and N. N. Dalvi. Foundations of probabilistic answers to queries. In *ACM SIGMOD International Conference on Management of Data*, 2005.

[25] J. S. Vitter and M. Wang. Approximate computation of multidimensional aggregates of sparse data using wavelets. In *ACM SIGMOD International Conference on Management of Data*, 1999.

[26] Q. Zhang, F. Li, and K. Yi. Finding frequent items in probabilistic data. In *ACM SIGMOD International Conference on Management of Data*, 2008.